# Nanoscale η-NiSi formation via ion irradiation of Si/Ni/Si

Nasrin Banu[1], Biswarup Satpati[2], Anjan Bhukta[3], B. N. Dev[1*]

[1]Department of Materials Science, Indian Association for the Cultivation of Science, 2A & 2B Raja S. C. Mullick Road, Jadavpur, Kolkata 700032, India.

[2]Surface Physics and Material Science Division, Saha Institute of Nuclear Physics, 1/AF, Bidhannagar, Kolkata 700064, India.

[3]Institute of Physics, Sachivalaya Marg, Bhubaneswar 751005, India

Abstract:

Nickel monosilicide (NiSi) has emerged as an excellent material of choice for source-drain contact applications below 45 nm node complementary metal-oxide-semiconductor (CMOS) technology. We have investigated the formation of nanoscale NiSi by ion irradiation of Si (~5 nm)/Ni(~15 nm)/Si, grown under ultrahigh vacuum environment. Irradiation was carried out at room temperature with 1 MeV $Si^+$ ions. X-ray diffraction (XRD) and transmission electron microscopy (TEM) were employed for analysis. With increasing ion fluence ion beam mixing occurs and more and more Si is incorporated into the Ni layer and this layer gets amorphized. At an even higher fluence a recrystallized uniform nickel monosilicide (η-NiSi) layer is formed. Several planar spacings of different Miller indices of η-NiSi have been observed in XRD and TEM. Additionally, an oscillatory amorphization and recrystallization has been observed in the substrate Si with increasing ion fluence. To our knowledge, this has never been observed in ion irradiation of bare Si in decades of work in this area. The oscillatory amorphization/recrystallization in Si is apparently Ni-induced. Irradiation displaces Ni and produces a distribution of Ni in amorphized Si. Irradiation at a higher fluence produces two recrystallized Si bands in amorphous Si with concomitant accumulation of Ni at the amorphous/crystalline interfaces. On further increase of irradiation fluence the recrystallized Si bands again pass through amorphization and recrystallization. The total thickness of recrystallized as well as amorphous Si shows an oscillatory behavior as a function of ion fluence.

## I. Introduction

As the device dimensions decrease in complementary metal-oxide-semiconductor (CMOS) technology with each new generation, new silicide source/drain contacts are developed to retain a low-resistivity silicide phase to meet the demands of scaling requirements [1,2,3]. Transition metal silicides and their nanostructures on silicon are of great current interest [4,5,6] as they are widely used in microelectronics industry. For nanoscale applications, among all transition metal silicides, particularly nickel monosilicide (NiSi) has emerged as an excellent material of choice for source-drain contact applications below 45 nm node CMOS technology [7,8]. However, there are major experimental challenges in growing uniform single phase nickel monosilicide on silicon. Various additive elements (impurities) have been used in the Ni layer for the investigation of the phase formation and morphological stability of nickel monosilicide films, which are grown by a thermal treatment. While some additive elements were found to retard agglomeration of the NiSi film, some other elements were found to retard the formation of $NiSi_2$. Both the problems could not be tackled with a single additive element [9]. Here we show that a uniform nanoscale single phase nickel monosilicide (NiSi) film can be grown by irradiating a pure Ni film, grown on Si in an ultraclean environment, with an energetic ion beam at room temperature.

As such energetic ion beams play a dominant role in silicon device fabrication. The impact of energetic ions on a solid material and their penetration into the material can modify the physical, chemical, electrical or magnetic properties of the solid material. The exposure of the material to the ion beams is known as ion irradiation. The incident ions collide with atoms in the solid, thereby displacing them from their lattice positions. When the incident ions lose all the energy via successive collisions with the atomic nuclei and electrons in the material, they stop at a depth and get incorporated into the material. This process is known as ion implantation. Ion implantation is heavily used for the fabrication of silicon devices [10].

The quality of thermally grown metal silicides can also be improved by ion irradiation [11]. As materials under ion irradiation undergo significant atomic rearrangement, atomic intermixing, alloying or new phase formation occurs at the interface between two materials during ion irradiation. Thus metal silicides can be formed by ion beam mixing when a metal layer, deposited on Si, is irradiated by energetic ions [12].

Here we present the formation of nanoscale nickel monosilicide (NiSi) by ion irradiation at room temperature. We also report on another novel phenomenon in the same system regarding the amorphization and recrystallization of the substrate silicon. Amorphization and recrystallization are fundamental processes and have been of significant interest since the beginning of ion implantation for the fabrication of silicon devices [13]. There have been many experiments of ion irradiation of bare Si substrates over several decades. According to the existing literature, at a given ion flux and given sample temperature, either amorphization or crystallization can occur and we are not aware of crystallization occurring at room temperature [13,14,15]. In our case of Si/Ni/Si, for a given flux and at room temperature, we observe an oscillatory amorphization and recrystallization in the Si substrate as a function of ion fluence.

The Ni layer in the Si(5 nm)/Ni(15 nm)/Si system is polycrystalline. Upon ion irradiation Si gets incorporated into the Ni layer and this Si-incorporated Ni becomes amorphized. Then at a higher fluence of irradiation this amorphous layer is recrystallized into the η-NiSi phase of nickel monosilicide.

## II. Experimental details

The sample was prepared in a molecular beam epitaxy (MBE) chamber under ultrahigh vacuum (UHV, ~$10^{-11}$ mbar) condition. First, a crystalline Si(c-Si) buffer-layer (~120 nm) was grown on a clean Si(001) wafer substrate at 700° C, the typical epitaxial growth temperature for Si on Si. (A buffer-layer is usually used to improve the quality of technologically important materials, epitaxially grown on them [16]). A Ni layer (~15 nm) was then grown on the buffer-Si layer at room temperature (RT). Finally an amorphous Si (a-Si) layer (~5 nm) was grown on Ni at RT. This was to prevent the oxidation of the Ni layer when the sample is taken out of the UHV growth chamber. Then the sample was removed from the UHV chamber. Before the ion irradiation experiments were carried out, presumably a native oxide of Si (~2 nm) has formed on the top surface.

Ion irradiation was carried out under high vacuum condition (mid $10^{-7}$ mbar) at RT with 1-MeV Si$^+$ ions in the fluence range of $1 \times 10^{14}$ - $3 \times 10^{17}$ ions/cm$^2$ with an average flux of $2.5 \times 10^{12}$ ions/cm$^2$.s. However, major changes that occur in the fluence range of $5 \times 10^{16}$-$3 \times 10^{17}$ ions/cm$^2$ will be predominantly discussed here. Following irradiation, the samples were investigated by X-ray diffraction (XRD) (with Cu K$_\alpha$ radiation), transmission electron microscopy (TEM) (with 300 keV electrons), scanning TEM (STEM) and energy dispersive X-ray (EDX) analysis.

## III. Results and discussion

### A. Formation of nickel monosilicide ($\eta$-NiSi)

XRD results obtained from the pristine sample as well as the irradiated samples are shown in Fig. 1. The (004) and the (002) (forbidden) reflections from the silicon substrate are seen. We also notice a Ni(111) peak from Fig. 1. The Ni layer has an average [111] surface normal orientation.

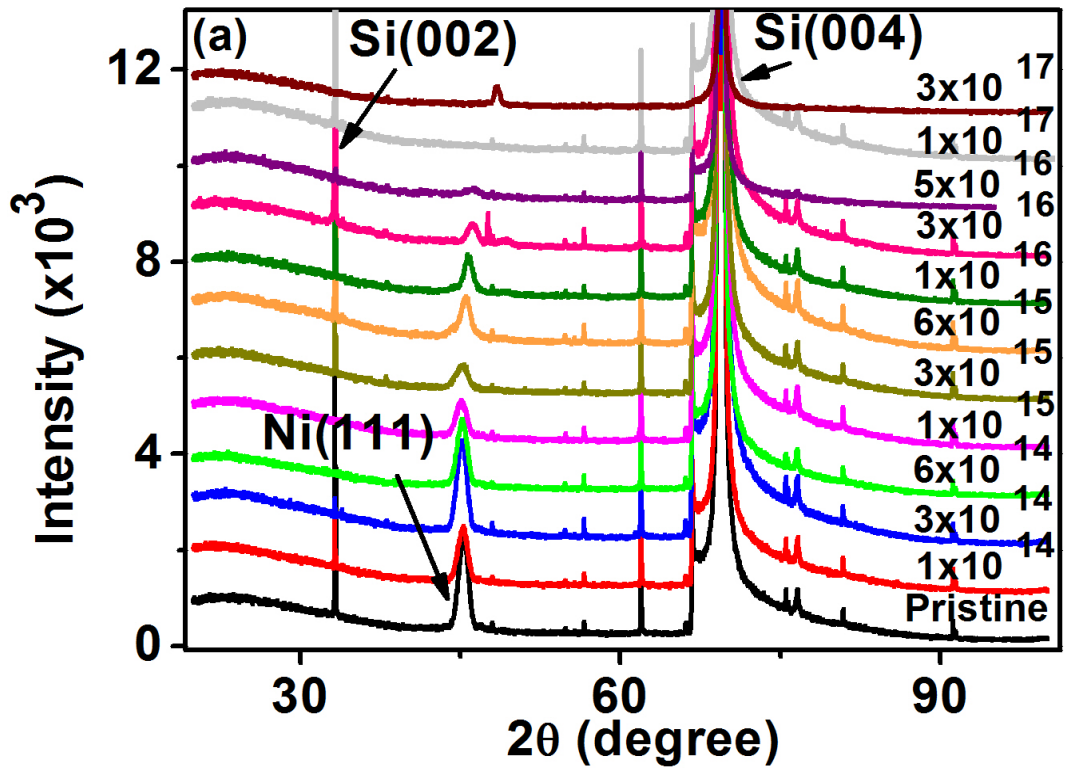

Fig. 1. XRD pattern from the pristine and the irradiated Si/Ni/Si samples, obtained with Cu K$_\alpha$ X-rays. Vertical shifts have been given to successive spectra for clarity.

The details about the uniformity of the Ni film and its polycrystalline nature are seen from TEM images. Cross-sectional TEM (XTEM) results from the pristine sample are shown in Fig. 2. A high resolution image from the Ni/Si interface region and the corresponding fast Fourier transform (FFT) pattern are shown in the inset (a) and (b), respectively. Below the Ni layer, the buffer-Si layer (~120 nm) region can be identified. The interface between the buffer-Si and substrate Si can also be identified (also in c). Fig. 2(d) shows the combined Si and Ni map, obtained by collecting Si L- and Ni M-X-rays in EDX, corresponding to the XTEM image in (c).

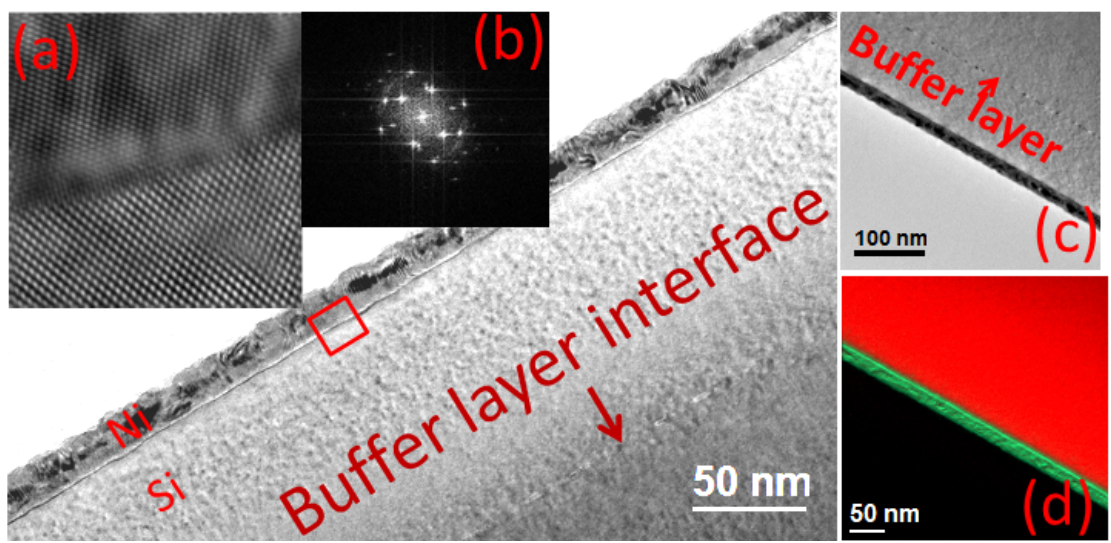

Fig. 2. XTEM image from the pristine sample. The interface between the buffer-Si layer and the substrate Si is shown by an arrow. (a) High resolution image from the Ni/Si interface region (boxed). (b) FFT pattern from the image in (a). (c) XTEM image and (d) the corresponding Ni map (green) and Si map (red), obtained by collecting Ni M- and Si L-X-ray fluorescence.

From Fig. 1 we notice that in ion irradiation with increasing ion fluence the Ni(111) peak gradually diminishes. At a fluence of $5 \times 10^{16}$ ions/cm$^2$, the Ni peak nearly vanishes and at $1 \times 10^{17}$ ions/cm$^2$, the Ni peak completely vanishes. Does it mean that Ni has been amorphized? As we will see later that the Ni layer is no more pure Ni and due to ion beam mixing a significant amount of Si has been incorporated into the Ni layer.

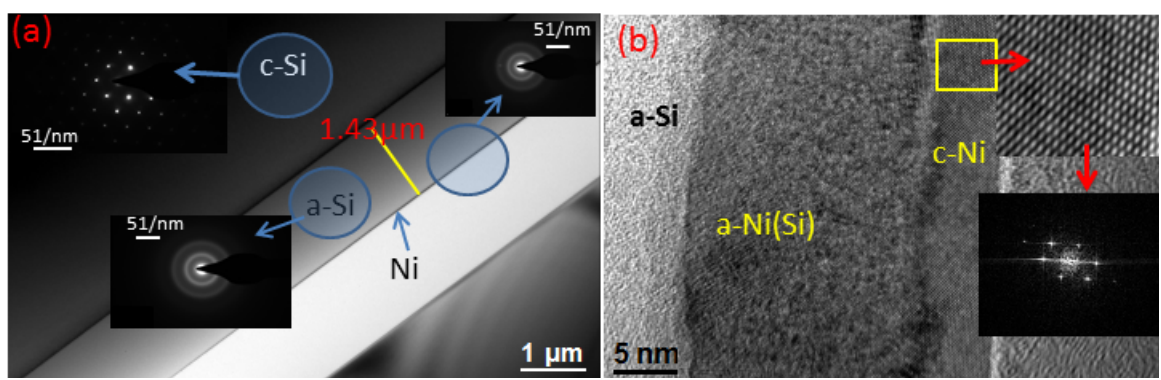

Fig. 3. XTEM images from the sample irradiated with $5 \times 10^{16}$ ions/cm$^2$ (a,b). (a) Diffraction patterns from different regions are shown. A uniform amorphous Si region of 1.43 μm thickness from the top has formed due to irradiation. (b) Ni, with incorporated Si [Ni(Si)] has also been amophized, except for a very thin crystalline layer at the top. Lattice image and the FFT from the crystalline region are seen.

Fig. 3 shows XTEM images from the sample irradiated at a fluence of $5 \times 10^{16}$ ions/cm$^2$, where the Ni(111) XRD peak nearly vanishes. In Fig. 3(a), we notice that Si has been amorphized up to a depth equivalent to the ion range, i.e. the depth ions could penetrate into the substrate

Si. The diffraction pattern (ring pattern) from this region indicates that it is amorphous Si (a-Si). Below this layer we see the diffraction pattern of crystalline Si (c-Si). The details about the Ni layer are seen in Fig. 3(b). From the substrate a-Si side, the Ni layer with incorporated Si has been amorphized [a-Ni(Si)]. A very thin layer (~ 5 nm) at the top still remains crystalline Ni (c-Ni). A high resolution lattice image from this region, along with its FFT pattern, is shown in the inset.

At an irradiation fluence of $1 \times 10^{17}$ ions/cm$^2$, we notice from Fig. 1 that the Ni peak has completely vanished. This means that the Ni layer, with incorporated Si, is completely amorphous. STEM high-angle annular dark field (HAADF) images for the fluences $1 \times 10^{17}$ and $2 \times 10^{17}$ ions/cm$^2$ and the corresponding elemental depth profiles of Si and Ni are shown in Fig. 4. In HAADF image a layer with higher atomic number elements looks brighter. The brightest layers in (a) and (c) are Ni. The regions in (a) with intermediate brightness are due to Ni accumulation at lower concentrations in these regions of Si, as will be seen later in direct elemental imaging. The dominant presence of Si in the Ni layer is clearly seen for the sample irradiated at a fluence of $1 \times 10^{17}$ ions/cm$^2$ (b). Si concentration in Ni further increases at an irradiation fluence of $2 \times 10^{17}$ ions/cm$^2$ (d), as indicated by the increasing ratio of Si counts in the Ni layer to that in the Si substrate.

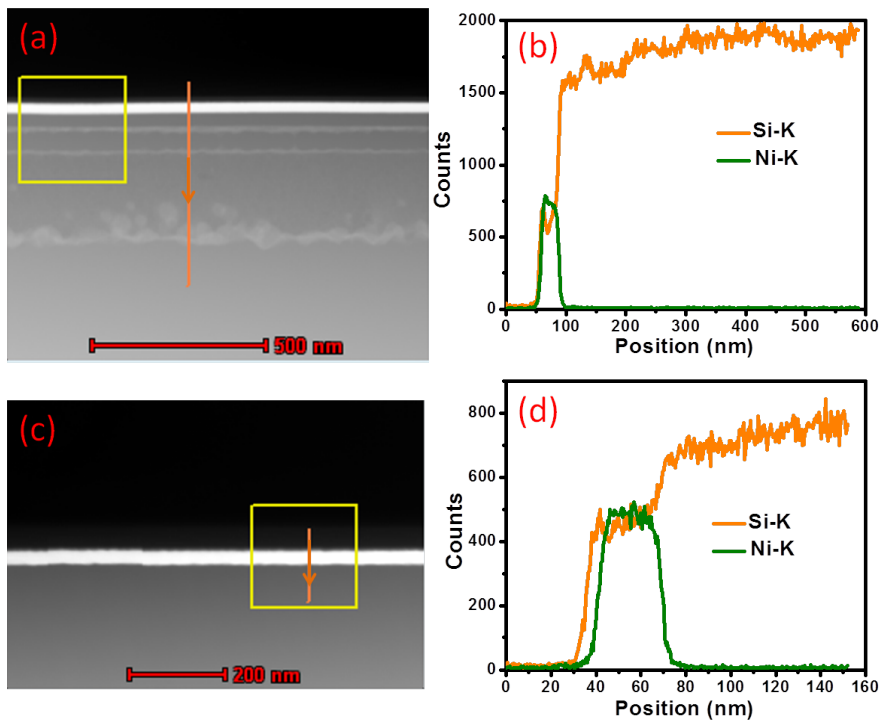

Fig. 4. STEM-HAADF XTEM images, (a) and (c), and the corresponding Ni and Si depth profiles, (b) and (d), along the marked lines in (a) and (c). (a-b): irradiation fluence is $1 \times 10^{17}$ ions/cm$^2$, (c-d): irradiation fluence is $2 \times 10^{17}$ ions/cm$^2$.

Si incorporation into Ni is also seen from the elemental map corresponding to the XTEM image shown in Fig. 5. In Fig. 5(a) the darkest layer is Ni. The bands marked '3' and '5' are recrystallized Si. We will discuss about this later. The Si map (red) in (b) reveals the presence of Si at the position of the Ni layer, i.e., within the Ni layer. The Ni layer (green) is seen in the Ni map (c) as well as in the combined Si and Ni map (d). In (c) and (d) Ni accumulation is also observed in two thin bands in Si. This will be explained later.

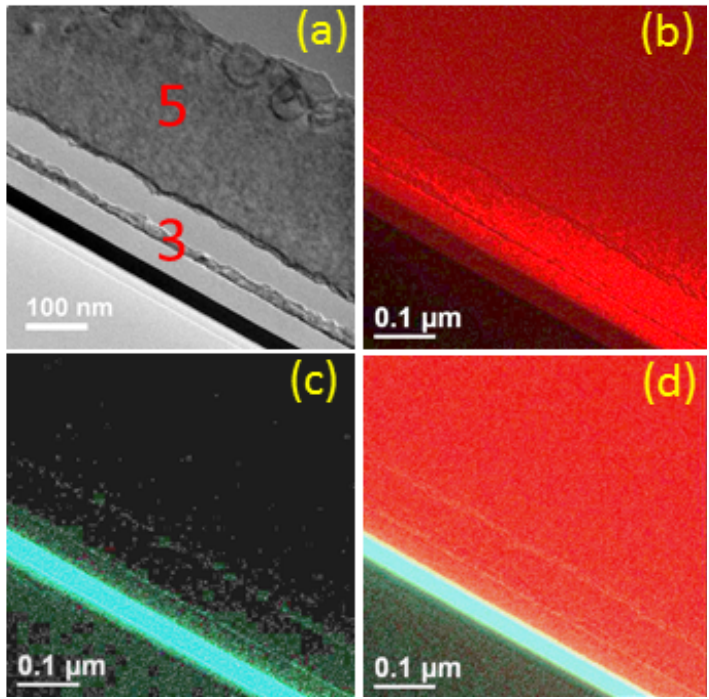

Fig. 5. (a-d) TEM results for the $1 \times 10^{17}$ ions/cm$^2$ fluence. (a) XTEM image. (b) Corresponding Si map (red), (c) Ni map (green) and (d) combined Si and Ni map. Ni accumulation at the top of the recrystallized Si layers (3 and 5) is seen.

Fig. 6 shows the results from the sample irradiated at $3 \times 10^{17}$ ions/cm$^2$. The XRD pattern from the Ni peak region is shown again (a) for the details. At this fluence, a new peak has appeared at a position considerably different from the Ni(111) peak position. This peak position corresponds to a planar spacing of 1.89 Å. Pure Ni does not have this planar spacing for any (hkl) Miller indices. As we have already seen incorporation of Si in Ni, this peak is expected to be from a phase of nickel silicide. The XTEM image is shown in Fig. 6(b). Layer '1' shown by the arrow is the Ni (rather Si-Incorporated Ni) layer. An image from this layer is shown at higher magnification in (c). The inset in (c) (left corner) shows a FFT pattern from the boxed region. In order to identify this phase we carried out diffraction pattern simulations for various Ni-Si compound phases. By comparing our simulated diffraction pattern with the FFT pattern in Fig. 6(c), we identify the material to be η-NiSi [17]. The simulated [18] diffraction pattern for [110] zone axis of η-NiSi is shown in the right corner inset of Fig.6(c). The close match between the FFT pattern and the simulated pattern is a good indication that the material is η-NiSi. However, we explore further details. A Fourier-filtered lattice image from the boxed region in (c) is shown in (d). We identify two planar spacings of 2.45 Å.

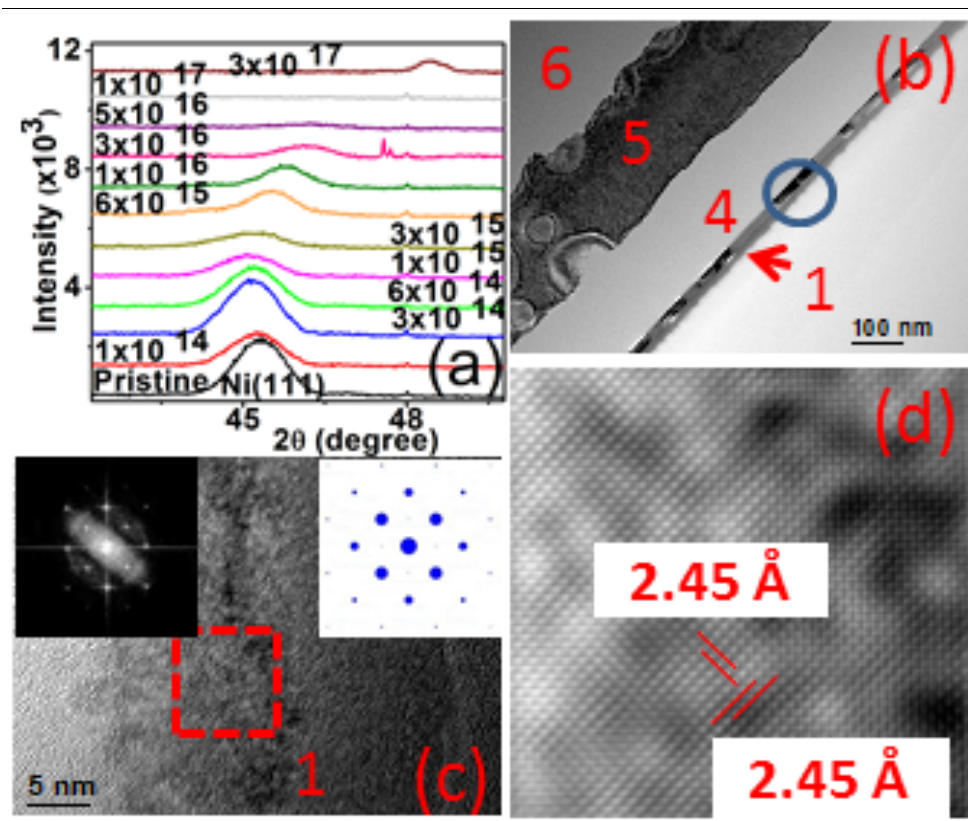

Fig. 6. (a) Evolution of the Ni(111) peak in XRD with fluence. Irradiation fluence is marked for each sample. (b–d) TEM results for a fluence of $3\times10^{17}$ ions/cm$^2$. (b) XTEM image. (c) Magnified image from the encircled region in (b), the FFT pattern (left inset) from the boxed region in (c) and a simulated pattern (right inset) for η-NiSi. (d) A high resolution image from the boxed region in (c) showing lattice planes.

Results of planar TEM investigations on this sample are shown in Fig. 7. A diffraction pattern from the encircled region in (a) is shown in (b). As there is a-Si underneath the NiSi layer, the ring pattern from a-Si is observed. They are marked. The diffraction spots from textured nanocrystalline η-NiSi are also marked. A high resolution image in (c) shows two crystallites. Two lattice images from the boxed regions in (c) are shown in (d) and (e). They show two planar spacings of 2.51 Å and 1.79 Å. Some planar spacings and the corresponding Miller indices (hkl) of η-NiSi are: 2.44 Å (301), 2.50 Å (113), 1.81 Å (411), 1.89 Å (303) [17]. While the first three planar spacings are observed in TEM, XRD shows the diffraction peak corresponding to the planar spacing of 1.89 Å. Nanocrystalline η-NiSi structures are seen in the plan view TEM image (Fig.7).

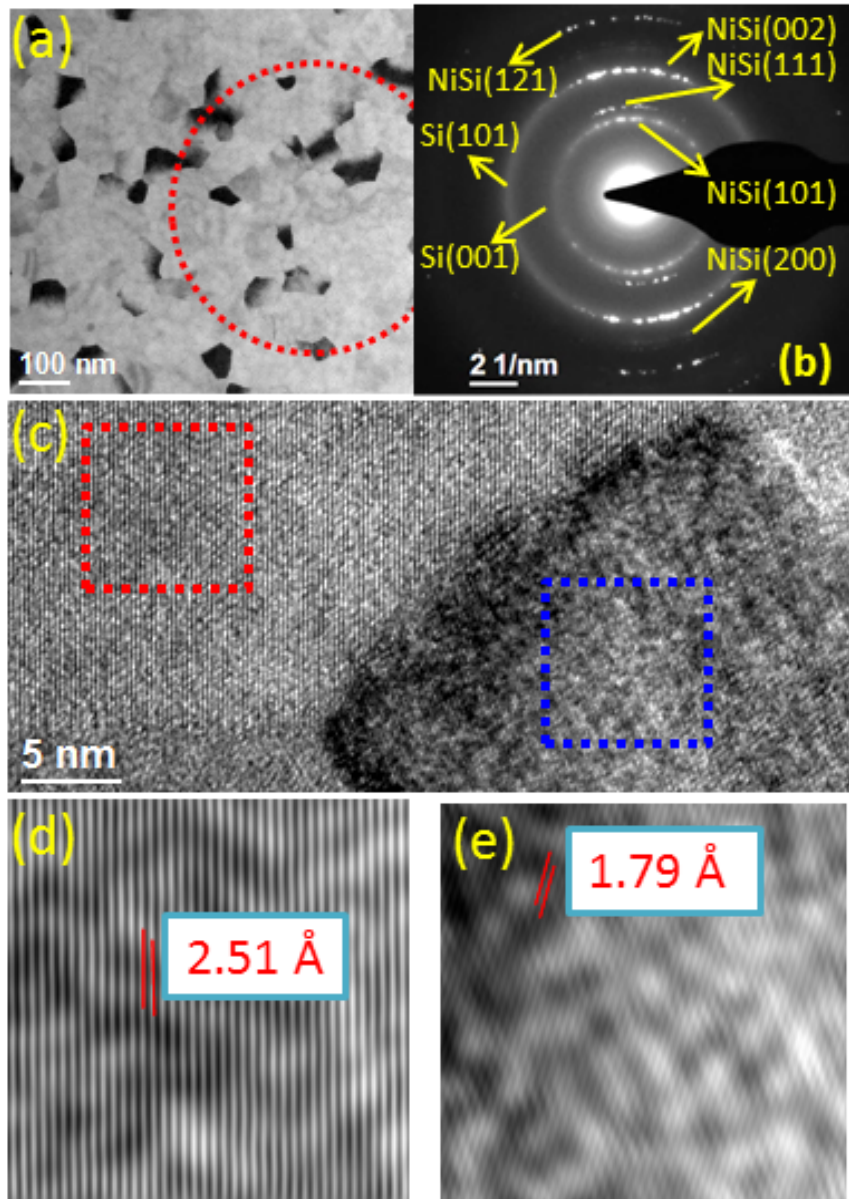

Fig. 7 (a) Planar TEM image. (b) Diffraction pattern from the encircled region in (a) with marked Miller indices of η-NiSi. (c) a higher magnification image showing differently oriented grains. (d) and (e) Fourier-filtered lattice images from the red and the blue boxed regions in (c), respectively.

The nickel monosilicide (NiSi) formed by ion irradiation at room temperature, as shown here, is free from the usual drawbacks, such as agglomeration and formation of $NiSi_2$, encountered in usual methods for the growth of NiSi. Additive elements are not required either. Clean contamination-free nickel monosilicide is formed by ion irradiation.

### B. Oscillatory amorphization and recrystallization of Si

Ion-beam-induced amorphization and recrystallization are fundamental processes in ion-solid interaction. These processes, especially in silicon, have been of significant interest since the beginning of ion implantation for the fabrication of silicon devices [13]. Review articles on this topic [13,15] throw light on an understanding of these phenomena. Ion irradiation through an amorphous/crystalline interface may stimulate recrystallization or amorphization depending on the sample temperature and ion beam parameters. At a constant ion flux, there is a substrate temperature $T_R$ (reversal temperature) such that, when $T > T_R$, ion irradiation produces epitaxial regrowth, whereas when $T < T_R$ the irradiation produces a layer-by-layer amorphization [14,15]. However, both amorphization and recrystallization do not usually occur at a given sample temperature. Moreover, we are not aware of any ion beam induced recrystallization occurring at RT. In our nanostructured Si(5 nm)/Ni(15 nm)/Si system we observe ion fluence dependent oscillatory amorphization and recrystallization in the substrate silicon at RT and at a constant ion flux.

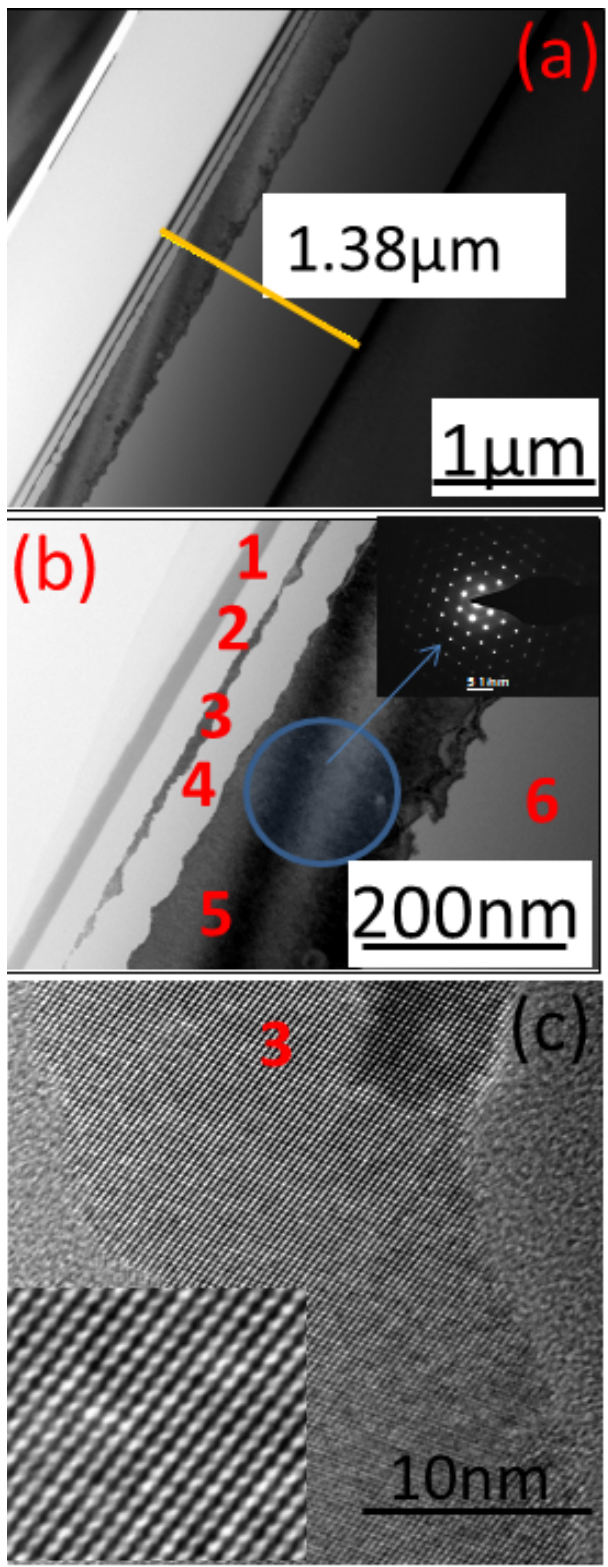

Fig. 8. XTEM images from the sample irradiated at $1 \times 10^{17}$ ions/cm$^2$. (a) Ion range is marked. (b) Higher magnification image showing different sub-layers in (a): 1 is Ni(Si), 2, 4 and 6 are a-Si, 3 and 5 are c-Si. C-Si diffraction pattern from layer 5 is shown in the inset of (b). (c) lattice image from layer 3 shows c-Si.

At an irradiation fluence of $5 \times 10^{16}$ ions/cm$^2$ ($\phi_1$) we observe the usual amorphization of Si up to a depth equivalent to the ion-range (Fig.3a). The diffraction patterns from different depths of the sample are shown. Beyond the ion-range Si is crystalline and within the range (~1.4 μm) Si has been amorphized. When the irradiation fluence is increased to $1 \times 10^{17}$ ions/cm$^2$ ($\phi_2$), two recrystallized (rc) bands of Si (Fig. 5a, bands 3 and 5) grow within the a-Si layer. Different bands or sub-layers are marked in Fig. 8(b). The diffraction pattern from the wider band (~350 nm, region 5) in Fig. 8(b) is that of c-Si. Also the crystalline lattice of Si is seen in Fig. 8(c) in the narrower band (~ 15 nm, region 3 in Fig.8(b)). Regions 2, 4 and 6 in Fig. 8(b) are a-Si. Fig. 9 shows the XTEM images from the sample irradiated at a fluence of $2 \times 10^{17}$ ions/cm$^2$ ($\phi_3$) and $3 \times 10^{17}$ ions/cm$^2$ ($\phi_4$). Here we notice that at $\phi_3$ the rc-Si bands (3 and 5) have been amorphized producing a continuous a-Si layer up to the depth equivalent to the ion-range, like the case of irradiation at $\phi_1$. At $\phi_4$ the Si in band 5 is recrystallized; however, there is no recrystallization of band 3.

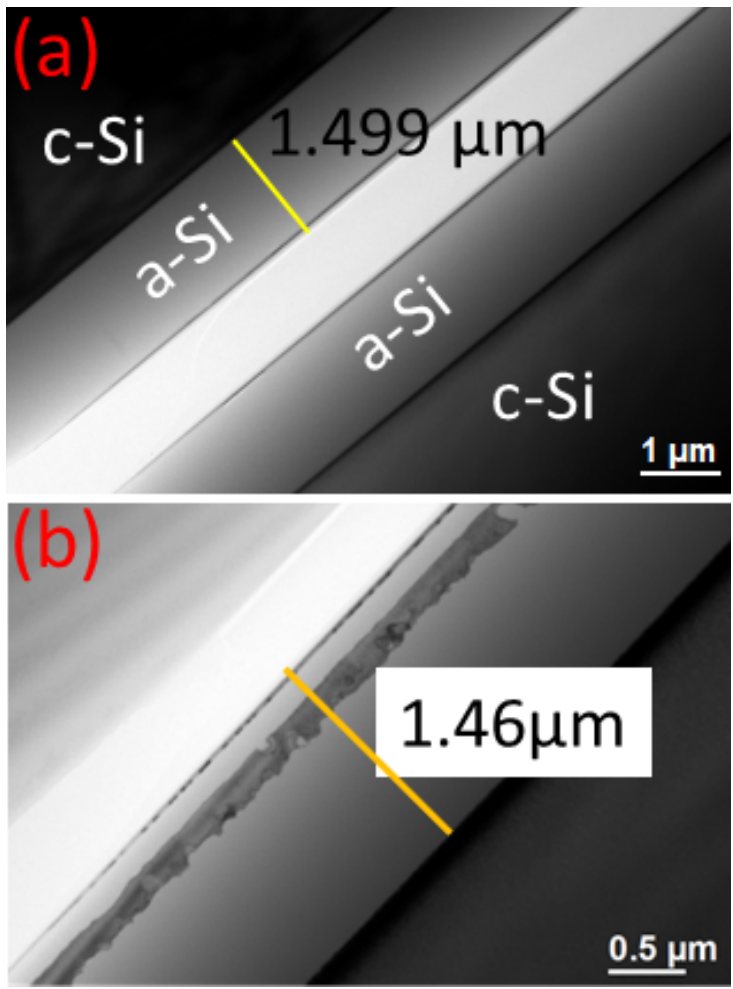

Fig. 9. XTEM images from the samples irradiated at a fluence of (a) $2 \times 10^{17}$ ions/cm$^2$ and (b) $3 \times 10^{17}$ ions/cm$^2$. In (a) two face-to-face bonded samples in TEM preparation are seen.

Let us discuss the oscillatory amorphization/recrystallization in Si by presenting the thickness of each layer as a function of fluence (Fig.10). "Range from TEM" is the distance from the top to the deepest a-Si/c-Si interface. Total amorphous thickness is the sum of thicknesses of all amorphous layers (2,4,6). The thickness of layer 6 closely follows the "total amorphous" curve, as the thicknesses of layer 2 and 4 are small. At $\phi_1$, as there are no rc-Si layers, 3 & 5, these two thicknesses are equal to zero. At $\phi_2$, both 3 & 5 rc-Si layers are present. In addition, a-Si layers 2 & 4 are present. That is why layer 6 is marginally smaller than the total amorphous thickness. With the increase of fluence, these thicknesses have an oscillatory behavior. Let us look at the thicknesses of the rc-Si layers 3 & 5. At $\phi_1$, their thicknesses are zero. At $\phi_2$, the thicknesses of the c-Si layers 3 & 5 increase. At $\phi_3$, both the values reduce to zero again. At $\phi_4$, layer 3 does not recrystallize; however, layer 5 is recrystallized. The thickness of layer 5 closely follows the total thickness of the rc-Si layers (3 and 5). Both the total thickness of the amorphous layers and the total thickness of the recrystallized layers show an oscillatory behavior as a function of ion fluence following an opposite trend.

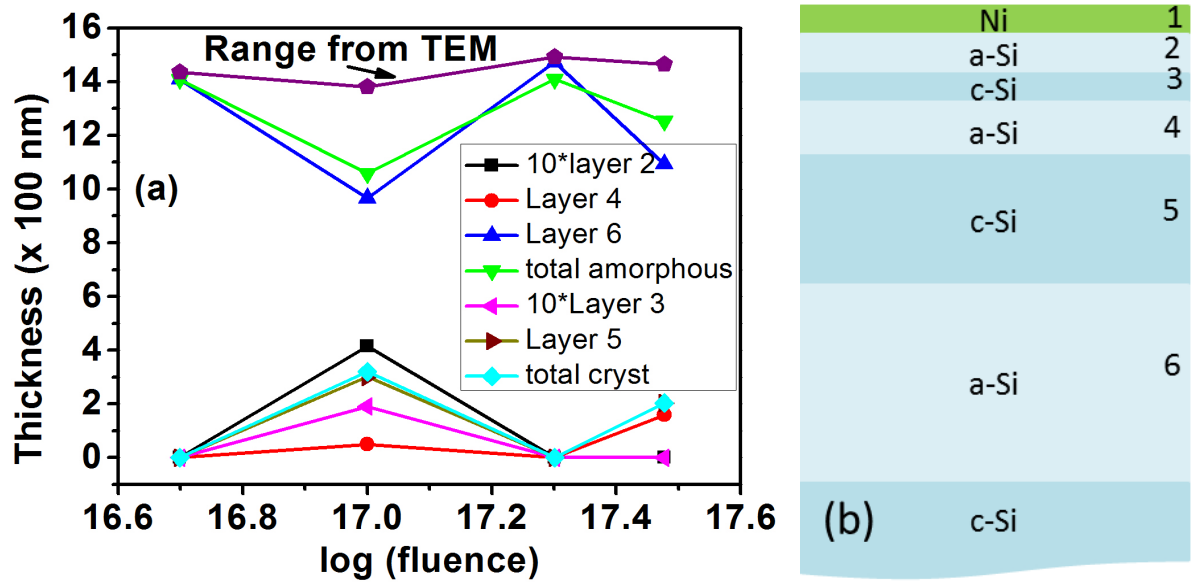

Fig. 10 (a) Oscillatory behaviour of thickness of different crystalline and amorphous layers of Si as a function of ion fluence. (b) The layer structure is schematically shown.

Let us explore the origin of this phenomenon. The case for $\phi_1$ is well understood. This amorphization occurs in ion implantation or irradiation. We investigate the case for $\phi_2$, when the rc-Si bands have formed. At $\phi_1$, when the continuous a-Si layer has formed, ion irradiation has caused displacement of Ni into the substrate Si. This Ni distribution is obtained by TRIM simulation [19] and is shown in Fig. 11(a). So, Ni is distributed in Si even beyond the buffer-Si/substrate-Si interface. A ballistic process is responsible for this. However, at higher ion fluences, enhanced diffusion and chemical processes can set in [20] and this can be responsible for recrystallization and Ni redistribution. Ion fluence dependent impurity migration, redistribution and chemical processes have been earlier investigated [21,22].

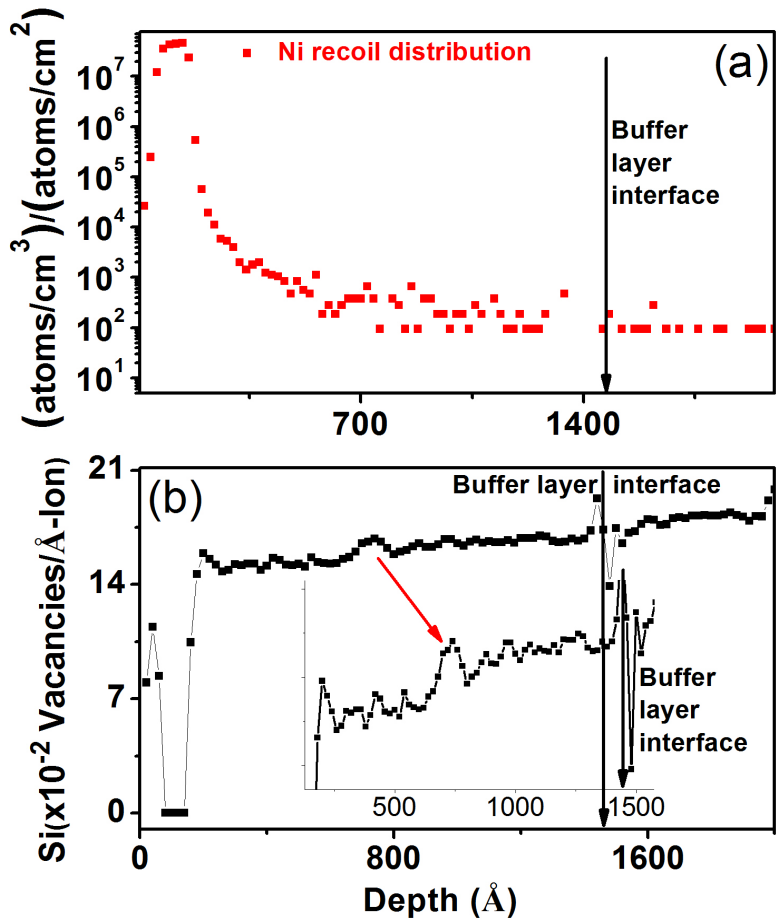

Fig. 11 TRIM simulation results for 1 MeV Si$^+$ ions incident on the Si(5 nm)/Ni(15 nm)/Si system. (a) Ni recoil distribution as a function of depth, (b) Si-vacancy depth distribution. The buffer-Si/substrate-Si interface is marked by a down arrow.

The presence of impurity atoms dispersed within an a-Si layer can dramatically affect the recrystallization process. Impurities which have high diffusivity have enough mobility to be redistributed at the advancing c/a interface during recrystallization. This modifies the initial impurity profile through the segregation towards the surface or interface [23]. In the recrystallization process of a-Si, impurity atoms are basically expelled from the recrystallized region and these impurity atoms accumulate at defect sites, such as interfaces. In fact this phenomenon, called gettering [24], is used to remove impurities from the region of interest for Si device fabrication. In Fig. 5(c),(d) two narrow bands of Ni accumulation within Si below the Ni layer are seen. The deeper band is at the buffer-Si/substrate-Si interface. Location of this interface is seen in Fig.2. This is where the deeper layer of Ni accumulation, at the top of the recrystallized Si band 5, takes place (see Fig. 5). What is the reason for Ni accumulation at the shallower band (at the top of recrystallized Si band 3)? TRIM simulation in Fig. 11(b) shows the Si vacancy distribution caused by ion irradiation for the Si/Ni/Si system. From the vacancy distribution (also shown in the inset) a peak, marked by an arrow, is seen. This vacancy peak happens to be at a depth coinciding with the shallow band of Ni accumulation. Recrystallization of Si band 3 and 5 expels Ni from these recrystallized Si bands into the two defected regions, namely, the buffer-Si/substrate-Si interface and the vacancy-rich region, where Ni is accumulated. TRIM simulation for a pure Si sample, i,e. without the Ni layer, does not show this vacancy peak. Here, the recrystallization process is induced by Ni. However, this is different from the standard metal-induced crystallization (MIC), which occurs only at elevated temperatures [25] and no ion irradiation is used. No oscillatory amorphization/recrystallization phenomenon in Si was observed earlier in ion irradiation.

## IV. Conclusion

Ion irradiation effects in a nanoscale Si/Ni/Si system have been investigated. Initially a buffer layer (~120 nm) of Si was epitaxially grown on a Si(001) substrate. A Ni layer (~15 nm) was deposited on the buffer Si layer and finally an amorphous Si layer (~ 5 nm) was deposited on Ni. Ion irradiation was carried out with 1 MeV $Si^+$ ions at room temperature and at a relatively low constant ion flux. With increasing ion fluence in irradiation, an increasing amount of Si is incorporated into the Ni layer and this layer is amorphized. At a higher fluence, this amorphous layer recrystallizes into the nickel monosilicide phase of η-NiSi. This method of ion irradiation for the formation of nickel monosilicide overcomes some usual difficulties for the growth of nickel monosilicide, which is a material of choice for applications below 45 nm node CMOS technology. While NiSi is formed in the Ni layer, another interesting oscillatory amorphization/recrystallization phenomenon occurs in the underlying Si substrate. For fluences of $5 \times 10^{16}$ ions/cm$^2$ onwards, the oscillatory amorphization/recrystallization phenomenon has been observed. A uniform amorphous Si (a-Si) layer is formed up to a depth equivalent to the ion range at a fluence of $5 \times 10^{16}$ ions/cm$^2$. Two narrow recrystallized Si bands form at a fluence of $1 \times 10^{17}$ ions/cm$^2$ within the a-Si layer. The recrystallized Si bands are amorphized again at a fluence of $2 \times 10^{17}$ ions/cm$^2$. At $3 \times 10^{17}$ ions/cm$^2$, again one recrystallized Si band reappears. This phenomenon has been identified to be Ni-induced. First, a distribution of ion-beam-induced displaced Ni is formed in the amorphized-Si region. Then at a higher fluence Si recrystallization occurs by expelling Ni from these recrystallized regions. These expelled Ni accumulates into two very thin (< 5 nm) defected regions – one at the interface of buffer-Si/Si-substrate and the other at a Ni-displacement-induced Si-vacancy-rich region. At higher fluences again Ni redistribution and Ni expulsion causes the oscillatory amorphization/recrystallization behavior. Thickness of the amorphous and the recrystallized Si regions as a function of ion fluence shows an oscillatory behavior.

## Acknowledgment

We thank the staff at the Ion Beam Laboratory, Institute of Physics, Bhubaneswar, India. Helpful discussion with Dr. Sumalay Roy is gratefully acknowledged. The work has been partially supported by the IBIQuS project (DAE OM No. 6/12/2009/BARC/R&D-I/50, Dated 01.4.2009). Nasrin Banu is supported by CSIR fellowship (09/080(0765)/2011-EMR-I).

*msbnd@iacs.res.in